\documentclass[12pt]{iopart}
\usepackage{setstack}
\begin{document}
\title{Dirty black holes: Quasinormal modes for ``squeezed'' horizons}
\author{A J M Medved, Damien Martin, and Matt Visser}
\address{School of Mathematical and Computing Sciences,
Victoria University of Wellington, PO Box 600, Wellington, New Zealand}
\eads{\mailto{joey.medved@mcs.vuw.ac.nz}, 
\mailto{damien.martin@mcs.vuw.ac.nz}, 
\mailto{matt.visser@mcs.vuw.ac.nz}}
\begin{abstract}
 
  We consider the quasinormal modes for a class of black hole
  spacetimes that, informally speaking, contain a closely ``squeezed''
  pair of horizons.  (This scenario, where the relevant observer is
  presumed to be ``trapped'' between the horizons, is operationally
  distinct from near-extremal black holes with an external observer.)
  It is shown, by analytical means, that the spacing of the
  quasinormal frequencies equals the surface gravity at the squeezed
  horizons.  Moreover, we can calculate the real part of these
  frequencies provided that the horizons are sufficiently close
  together (but not necessarily degenerate or even ``nearly
  degenerate'').  The novelty of our analysis (which extends a
  model-specific treatment by Cardoso and Lemos) is that we consider
  ``dirty'' black holes; that is, the observable portion of the
  (static and spherically symmetric) spacetime is allowed to contain
  an arbitrary distribution of matter.

\vskip 0.50cm
\noindent
  Dated:  29 October 2003; \LaTeX-ed \today
\\
arXiv: gr-qc/0310097
\end{abstract}

\maketitle

\def\d{{\mathrm{d}}}
\def\be{\begin{equation}}
\def\ee{\end{equation}}
\def\Im{\hbox{Im}}
\def\Re{\hbox{Re}}
\def\sech{\hbox{sech}}
\def\max{\hbox{max}}
\def\min{\hbox{min}}
\def\half{{1\over2}}
\def\define{\equiv}
\def\Nordstrom{Nordstr\"om}
\def\paragraph#1{ {\noindent{\bf #1}\quad}}
\def\phase{{\,\hbox{phase}\,}}

\def\ns#1{_{\mathrm{#1}}}

\section{Introduction}

The perturbations of the spacetime outside of a black hole horizon are
expected to be radiated away, at late times, with a discrete set of
complex-valued frequencies; the so-called {\it quasinormal mode}
frequencies of a black hole~\cite{Pr,XX3,Rev}.  Knowledge of these
modes should have particular importance in gravitational-wave
astronomy and, in a more speculative scenario, may even provide
insight into the very essence of black hole entropy (which still lacks
a convincing statistical explanation).

With regard to the latter motivation, an interesting proposal has been
put forth by Hod~\cite{Hod}.  On the basis of Bohr's correspondence
principle, Hod has suggested that, in the asymptotic limit of a
``highly damped'' black hole,~\footnote{By highly damped, it is meant
  that the imaginary portion of the mode frequency has become very
  large. This connection follows from the imaginary part being a
  measure of the inverse relaxation time of a radiating black hole.}
the real part of the quasinormal frequency should represent a
characteristic (transition) frequency for the black hole itself.
Moreover, it was then argued that the value of this special frequency
could be used as a means for uniquely fixing the level spacing of the
black hole area spectrum.  (The concept of a uniformly spaced area
spectrum for black holes was first advocated by
Bekenstein~\cite{Bek}.)

To help flesh out these somewhat esoteric statements, let us consider
the asymptotic behavior of the quasinormal modes for a Schwarzschild
black hole. In the case of scalar or gravitational perturbations,
these asymptotic frequencies are known to take the following form
\cite{Nol,And}:
\begin{equation}
k_{qnm}(n) =  {1\over 4m}\left[i\left(n+{1\over 2}\right)+{\ln 3\over
 2\pi}\right]   + {\cal O} [n^{-1/2}] 
\qquad {\rm  as} \qquad  n\rightarrow\infty \;,
\label{0}
\end{equation}
where $n$ is the quantum number that labels the frequencies and $m$ is
the black hole mass.~\footnote{Here and throughout, all fundamental
  constants are fixed to unity and a four-dimensional spacetime is
  presumed.}  Note that the asymptotic spacing, or the ``gap'', is
just the Schwarzschild surface gravity ($\kappa_s=1/4m$).  According
to Hod~\cite{Hod}, one can identify the real part of the subleading
term, or the ``offset'', as the ``transition frequency'' $\omega$ so
that $\delta m=\omega=\kappa_s\ln 3/2\pi\;$.  Then, given the
Schwarzschild area formula or $A=16\pi m^2\;$, it immediately follows
that $\;\delta A=32 \pi m \,\delta m=4\ln 3\;$ equals the spacing
between adjacent levels in the area spectrum.  It is an interesting
consequence that, with this particular spacing, $\exp(S)$ takes on an
integer value (where $S=A/4$ is the black hole entropy) in compliance
with statistical expectations.

The above considerations have helped trigger a recent surge of
research activity in the realm of quasinormal modes.  For instance,
there has been some discussion on linking the intriguing appearance of
$\ln 3$ with the gauge group of loop quantum gravity ({\it e.g.}
\cite{Dre,Kun}).  Furthermore, there has also been a considerable
amount of work on evaluating the quasinormal frequencies for various
black hole models; both numerically ({\it e.g.}~\cite{BK,GA}) and
analytically ({\it e.g.}~\cite{Motl,MN,Neitzke}).  For a more thorough
list of references, see~\cite{MMV} and~\cite{Cardoso2}.)

It is worth pointing out that most of these frequency calculations
have been highly model specific.  (For a recent status report and
summary, see~\cite{Cardoso2}.) On the other hand, it is the opinion of
the current authors that significant progress will almost certainly
require more general considerations. This perspective has prompted a
recent study~\cite{MMV} in which we investigated the asymptotic mode
behavior for a very general class of spacetimes: that of ``dirty''
black holes.  These are static, spherically symmetric but otherwise
{\it generic} spacetimes for which there is a central black hole
surrounded by arbitrary matter fields~\cite{DBH,DBH2,DBH3}.  (Keep in
mind that such geometries also allow for multi-horizon scenarios and
the presence of a cosmological horizon.)

Although our prior treatment~\cite{MMV} confirmed that the gap going
as the surface gravity is a generic result, it was unable to say
anything about the real part of the offset. In the current paper,
however, we are able to rectify this omission for a certain class of
dirty black hole spacetimes.  More specifically, we will now focus on
a scenario of ``squeezed'' horizons. That is to say, the observable
part of the spacetime is bounded by a pair of horizons which are, in a
well-defined sense, regarded as being ``close''. (This should not be
confused with the case of a nearly extremal black hole where the
observer is, most typically, regarded as residing in the exterior
portion of the spacetime.)  For this class of spacetimes, it will be
shown that the offset can, in fact, be calculated {\it exactly} in the
squeezing limit. This will be accomplished by casting the (generic)
``scattering potential'' into a recognizable, Poschl--Teller
form~\cite{PT}.  Notably, the same basic technique has been used by
Cardoso and Lemos~\cite{CL} for the specific case of a
Schwarzschild--de Sitter (``squeezed'') geometry; nevertheless, we are
now able to demonstrate that their results have much more general
applicability.

The remainder of the paper is arranged as follows.  In the following
section, we present the basic formalism and set up conventions.  The
third section contains the main analysis; that is, the calculation of
the quasinormal mode frequencies as previously discussed.  The main
body of the paper ends with a brief summary and discussion.  There is
also an appendix where the scattering potential is rigorously derived
for axial gravitational perturbations in a dirty black hole background
(the scalar-perturbation calculation is documented in~\cite{MMV}).

\section{The formal preliminaries} 
\label{S:II}

Let us begin by introducing our so-called {\it dirty} black hole
spacetime. That is, a spherically symmetric and static --- but
otherwise general --- black hole geometry (alternatively, the geometry
of a central black hole surrounded by arbitrary matter fields).
Without loss of generality, the metric for such a spacetime can always
be expressed as ({\it e.g.},~\cite{DBH,wormhole})
\begin{equation}
\d s^2 = - e^{-2\phi(r)}\; \left[1-{2m(r)\over r}\right] \d t^2 
+ \left[1-{2m(r)\over r}\right]^{-1} \d r^2 + r^2 \d \Omega^2,
\label{1}
\end{equation}
where $\phi(r)$ and $m(r)$ are model-dependent functions (related to
the Morris--Thorne ``redshift'' and ``shape'' functions,
respectively~\cite{Morris-Thorne}).  Keep in mind that $m(r)$ retains
its usual interpretation as a mass parameter, whereas $\phi(r)$
measures any deviation (due to matter fields) from the ``Schwarzschild
form'' of $g^{-1}_{rr}=-g_{tt}\;$.

It often proves useful to have the metric expressed in terms of a
generalized tortoise coordinate,
\begin{equation}
{\d r_*\over\d r} = { e^{\phi(r)} \over 1-{2m(r)\over  r} }\;,
\label{2}
\end{equation}
so that the line element (\ref{1}) can now be written as
\begin{equation}
\d s^2 = e^{-2\phi(r)}\; \left(1-{2m(r)\over r}\right) 
\left[ - \d t^2 + \d r_*^2\right] + r^2 \d \Omega^2 \;.
\label{3}
\end{equation}

Note that our formalism does, in its full generality, allow for
spacetimes with more than one causal horizon; one of which may be a
cosmological horizon (if the spacetime is asymptotically de Sitter).
For any such horizon --- that is, any radius $r=r_h$ which satisfies
$r_h=2m(r_h)$ --- one can calculate the surface gravity via standard
methods~\cite{DBH}.  More specifically, for a black hole (event)
horizon,
\begin{eqnarray}
\kappa_h  &=&
{1\over 2}\left.
{\d \over\d r}  \left[ e^{-\phi(r)} \; \left(1-{2m(r)\over r} \right) \right] 
\right|_{r_h} \label{4}
\\ &=&
{1\over 2 r_h} e^{-\phi(r_h)} \left[1-8\pi\rho(r_h)\; r_h^2\right]\;,
\nonumber
\end{eqnarray}
where $\rho(r)$ is the energy density. In contrast, one would take the
negative of the above expression to obtain the surface gravity at a
cosmological horizon~\cite{GH}.  In this case, the non-negativity of
surface gravity immediately implies a lower bound on the radius of any
cosmological horizon, $r_c$, as a function of the energy density;
namely, $r_c^2\geq [8\pi\rho(r_c)]^{-1}\;$. This inequality can be
viewed as a generalization of the Nariai bound~\cite{Nai}; that is,
the upper limit on the mass of a black hole in a Schwarzschild--de
Sitter [Kottler] geometry.

Let us now assume the existence of two non-degenerate horizons; say,
$r_a$ and $r_b$ with $r_b>r_a\;$ and no other horizons in between.
(There can, however, still be other horizons in the spacetime, but
these would be operationally irrelevant to our treatment --- see
below.)  When considering this particular scenario (especially
relevant to Section 3), we will {\it always} view matters from the
perspective of an observer who is ``trapped'' between the two
horizons. Hence, considerations can safely be restricted to the
submanifold defined by $r_a\leq r \leq r_b \;$.  [It is useful to
remember that, for this setup, the region $r\in\left(r_a,r_b\right)$
maps into the region $r_*\in\left(-\infty,+\infty\right)$.]  Under
these circumstances, the metric components of interest can always be
expressed as
\begin{equation}
g^{-1}_{rr}=1-{2m(r)\over r}={\left(r-r_a\right)\left(r_b-r\right)\over
r^2} \; h(r) 
\label{5}
\end{equation}
and
\begin{equation}
\left|g_{tt}\right|=e^{-2\phi(r)}\left(1-{2m(r)\over r}\right)
=e^{-2\phi(r)}{\left(r-r_a\right)\left(r_b-r\right)\over r^2}
\; h(r) \; ,
\label{6}
\end{equation}
where $h(r)$ is some dimensionless function that is well behaved,
regular and positive throughout the observable portion of the
manifold. Note that one can now write the mass as
$2m(r)/r=1-(r-r_a)(r_b-r)h(r)/r^2\;$, indicating that $h(r)$ can be
physically interpreted as (a dimensionless measure of) the energy
density of the extraneous or ``dirty'' matter.

For future convenience, let us re-express some earlier formalism
directly in terms of the function $h(r)$ or, rather, the modified form
(for calculational simplicity) $H(r)\equiv e^{-\phi(r)}h(r)/r^2\;$.
The defining expression for the generalized tortoise coordinate
(\ref{2}) now becomes
\begin{equation}
{\d r_*\over\d r} =  \left[(r-r_a)(r_b-r)H(r)\right]^{-1}\;;
\label{7}
\end{equation}
whereas the surface gravity (\ref{4}) can be re-evaluated to give (for
the inner and outer horizon, respectively)
\begin{equation}
\kappa_a={1\over 2}\left(r_b-r_a\right)H(r_a)\;,
\label{8}
\end{equation}
\begin{equation}
\kappa_b={1\over 2}\left(r_b-r_a\right)H(r_b)\;.
\label{9}
\end{equation}

Since the primary interest of this paper is the calculation of
quasinormal mode frequencies, let us remind the reader of some
pertinent points.  One considers small perturbations of the spacetime
outside of the relevant horizon(s).  As is well known in the
Schwarzschild case, such considerations ultimately lead to a
one-dimensional Schrodinger-like equation~\cite{RW}; that is,
\begin{equation}
{\d^2\over \d r_*^2} \psi - V[r(r_*)] \, \psi = - k^2 \psi\;,
\label{10}
\end{equation}
where $\psi=\psi\left[r(r_*)\right]$ describes the radial behavior of
the (massless) perturbation field, $k$ is the frequency and, for a
scalar (spin $j=0$) field in particular, the ``scattering potential''
is found to be
\begin{equation}
V(r) = \left(1-{2m\over r}\right)
\left[{\ell(\ell+1)\over r^2} + {2m\over r^3}\right]\;,
\label{11}
\end{equation}
with $\ell$ representing the orbital angular momentum
($\ell=0,1,2,...$).  In the case of higher-spin ($j>0$) fields, there
is an often quoted generalization of this result~\cite{Zer} whereby
one makes the replacement $2m\rightarrow 2m(1-j^2)$ in the numerator
of the right-most term.  However, it is certain that this
generalization cannot in all circumstances be correct; see the
appendix for further discussion.

We have, quite recently~\cite{MMV}, generalized the
scalar-perturbation potential to the generic spacetime described by
equations (\ref{1}) and (\ref{3}).  The one-dimensional
Schrodinger-like form (\ref{10}) does indeed persist, although the
scattering potential becomes somewhat more complicated.  More
precisely,
\begin{eqnarray}
V(r) = 
e^{-2\phi(r)}\left(1-{2m(r)\over r}\right) \label{12}
\\  \quad\quad\quad
\times
\left[{\ell(\ell+1)\over r^2}+ 2{m(r)\over r^3}
-2{m^{\prime}(r)\over r^2} 
 -\left(1-{2m(r)\over r}\right) {\phi^{\prime}(r) \over r}
\right] \;,
\nonumber
\end{eqnarray}
where a prime denotes a derivative with respect to $r$.  It is unclear
if there is a simple way to generalize this result to arbitrary values
of spin, $j$.  For instance, the form of the potential for the $j=2$
case of axial gravitational perturbations (which we derive in the
appendix) could not have easily been anticipated.  (That is, one could
not just make the type of naive replacement that was discussed above.)
Nonetheless, we propose that the precise dependence of the potential
on $j$ is, qualitatively speaking, inconsequential to the
prior~\cite{MMV} or the current treatment.

The quasinormal modes can be regarded as the {\it complex} frequency
solutions of equation (\ref{10}) when ``radiation boundary
conditions'' are imposed at the boundaries of the observable
spacetime~\cite{Pr,Rev}.  (Complex values of $k$ are, in fact,
necessitated by having a ``mostly positive'' potential.)  With regard
to the case of a Schwarzschild black hole, the quasinormal mode
frequencies are known to be labeled by a discrete quantum
number~\cite{XX3} and have a well-known asymptotic form, exhibited in
equation (\ref{0}) and amply confirmed both numerically~\cite{Nol,And}
and analytically~\cite{Motl,MN}.

More generally, we have recently shown that, up to the leading
asymptotic order, the Schwarzschild form for the mode frequencies is,
indeed, a generic feature of black hole spacetimes~\cite{MMV}.  More
to the point, the imaginary spacing between the modes (or the ``gap'')
is, in the asymptotic limit, exactly equal to the relevant surface
gravity under quite generic circumstances.  Also in the cited work, we
have extended considerations to an observer who is trapped between two
horizons (precisely, the scenario discussed earlier in this section)
and argued for the following generalization:~\footnote{The same
  expression has also been deduced, by independent means, for the
  specific model of a Schwarzschild--de Sitter black
  hole~\cite{Suneeta}. However, for a contrary opinion,
  see~\cite{PAD}.}
\begin{equation}
k_{qnm}(n) = i\left[n_a\,\kappa_a+n_b\,\kappa_b\right] + O[1]   
\qquad {\rm when} 
\qquad n_a,\,n_b \rightarrow\infty 
\;.
\label{14}
\end{equation}
Let us re-emphasize, however, that our prior approach~\cite{MMV} ---
which utilized the first Born approximation to the scattering
amplitude --- could not say anything about the real part of the
order-unity term or the ``offset''.

\section{The main analysis} 
\label{S:III}

We will now proceed to show that, for a certain class of observers, an
{\it exact} calculation of the offset is still ``generically''
possible.\footnote{By generic, we mean that the spacetime can be
  arbitrarily dirty.}  To elaborate, we will return to our scenario of
an observer trapped between horizons at $r_a$ and $r_b$, and then
consider the ``squeezed'' horizon limit of $r_a \rightarrow r_b$ (or
{\it vice versa}).  [This is, of course, just the degenerate or
extremal horizon limit; nonetheless, we have avoided using such
terminology, as the usual inference is an observer exterior to both
horizons ({\it e.g.}, an observer outside of a Reissner--Nordstr\"om
black hole).]  Note that the following analysis generalizes a prior
work, specific to the case of a Schwarzschild--de Sitter spacetime, by
Cardoso and Lemos~\cite{CL}.

Let us begin here by establishing what is exactly meant by a spacetime
with squeezed horizons.  For the duration, the horizons will be
regarded as sufficiently close so that $\Delta\equiv
\left(r_b-r_a\right)/r_a \ll 1\;$.  Given the extent of the relevant
manifold ($r_a\leq r\leq r_b\;$), it immediately follows that, up to
corrections of the order $\Delta\;$, $r_a\sim r \sim r_b\;$.

In view of the above, it is clear that much of the prior formalism
will simplify. For instance, let us recall equation (\ref{7}) for the
tortoise coordinate. Up to corrections of the relative order
$\Delta\;$ (always signified by $\sim$), we can write
\begin{equation}
{\d r_*\over\d r} \sim  \left[(r-r_a)(r_b-r)H(r_a)\right]^{-1}\;.
\label{15}
\end{equation}
To justify this, recall that $H(r)=e^{-\phi(r)}h(r)/r^2$ is a regular,
well-defined and positive function throughout the observable region
[{\it cf}, equations (\ref{5}) and (\ref{6})].  We can then safely
approximate
\begin{equation}
H(r_a) \sim H(r) \sim H(r_b)
\end{equation}
in the region of interest $r\in(r_a,r_b)$. In particular
\begin{equation}
\kappa_a \sim \kappa_b
\end{equation}
up to corrections of higher order in $\Delta$.  The approximate
relation for the tortoise coordinate can now readily be integrated to
give
\begin{equation}
r_*\sim {1\over \left(r_b-r_a\right)H(r_a)}\ln\left[{r-r_a\over
r_b-r}\right] 
=
{1\over 2\kappa_a}\ln\left[{r-r_a\over r_b-r}\right] 
\;,
\label{16}
\end{equation}
which can then  be inverted to yield
\begin{equation}
r\sim {r_a+r_b\exp\left[2\kappa_a r_*\right]\over
1+\exp\left[2\kappa_a r_*\right]}
={r_a+r_b\over 2}+{r_b-r_a\over 2}\tanh[\kappa_a r_*]
\;.
\label{17}
\end{equation}
Applying equation (\ref{8}), we can directly substitute the above
relation for $r=r(r_*)$ into $(r-r_a)(r_b-r)H(r_a)$ and obtain (after
some manipulations)
\begin{equation}
 \left(r-r_a\right)\left(r_b-r\right)H(r_a)
\sim {\kappa_a\left(r_b-r_a\right) 
\over 2\cosh^2 \left[\kappa_a r_*\right]}\;.
\label{19}
\end{equation}
But, given that $H(r)\sim H(r_a)$ throughout the relevant manifold,
this also means that
\begin{equation}
 e^{-\phi(r)}\left(1-{2m(r)\over r}\right)=
 \left(r-r_a\right)\left(r_b-r\right)H(r)
\sim {\kappa_a\left(r_b-r_a\right) 
\over 2\cosh^2 \left[\kappa_a r_*\right]}\;.
\label{20}
\end{equation}
It is useful to note that the left-hand side is essentially the
time-time component of the metric [{\it cf}, equation (\ref{6})], up
to a ``distortion'' which can be regarded as a constant factor in the
regime of interest.

The above result is, basically, all that is needed to ascertain the
quasinormal modes. To demonstrate this, we will initially concentrate
on the simplest case of a scalar perturbation and then comment on
general values of $j$ below.  To begin here, let us first call upon
the generic form of the scattering potential for scalar fields
(\ref{12}).  Conveniently, one can always re-express this potential in
the following compact form~\cite{MMV}:
\begin{equation}
V(r) = 
e^{-2\phi(r)}\left(1-{2m(r)\over r}\right)
{\ell(\ell+1)\over r^2} +{1\over r}\left(\partial^2_{r_*}r\right)\;.
\label{21}
\end{equation}
Evaluating the partial derivatives with the help of equation
(\ref{7}), we find that
\begin{eqnarray}
V(r) = 
\left(r-r_a\right)\left(r_b-r\right)H(r)
\label{22} \\ 
\quad\quad
\times
\left[e^{-\phi(r)}
{\ell(\ell+1)\over r^2} +{\left(r-r_a\right)\left(r_b-r\right)
H^{\prime}(r)\over r}
+{\left(r_a+r_b-2r\right)H(r)\over r}
\right]\;.
\nonumber
\end{eqnarray}
So far, everything is exact; however, applying the ``squeezed
approximation'' ($\Delta\ll 1\;$) and equation (\ref{20}) in
particular, we then get
\begin{equation}
V(r_*)\sim e^{-\phi(r_a)}{\kappa_a\left(r_b-r_a\right)\over
2\cosh^2\left[\kappa_a r_*\right]}{\ell(\ell+1)\over r^2_a}
=
{\kappa_a^2\over h(r_a)} \; 
{\ell(\ell+1)\over\cosh^2\left[\kappa_a r_*\right]}.
\label{23}
\end{equation}
To put it another way,
\begin{equation}
V(r_*)\sim {V_{j=0,\ell}\over \cosh^2\left[\kappa_a r_*\right]}\;,
\label{24}
\end{equation}
where $V_{0,\ell}$ is a model-dependent quantity (in principle, always
calculable) that is constant for any fixed value of $\ell$.

Significantly, equation (\ref{24}) can be identified as the
Poschl--Teller potential~\cite{PT}; a form of potential for which the
one-dimensional scattering equation (\ref{10}) can readily be solved.
Choosing appropriate plane-wave boundary conditions for the
perturbation field, $\; \Psi\sim \exp[ik(t\pm r_*)]$ as
$r_*\rightarrow\mp\infty\;$, one finds that~\cite{FM}
\begin{equation}
k_{qnm}\sim\kappa_a \left[i\left(n+{1\over 2}\right)
+\sqrt{{V_{0,\ell}\over\kappa_a^2}-{1\over 4}}\;\right]
\qquad {\rm where}\qquad n=0,1,2,... \;.
\label{25}
\end{equation}
This result generalizes equation (20) in reference~\cite{CL}, which
was specific to the case of a (``squeezed'') Schwarzschild--de Sitter
black hole.  It is easily confirmed that our form of the potential
(\ref{23}) reduces to theirs for this special model.

Take note of the gap or spacing between the levels at large $n$; this
is precisely $\kappa_a\sim\kappa_b\;$, an outcome which agrees with
our prior findings [{\it cf}, equation (\ref{14})].  Moreover, the
offset can now be trivially extracted:
\begin{equation}
i{\kappa_a\over 2}
+\sqrt{V_{0,\ell}-{\kappa_a^2\over 4}}
=
i{\kappa_a\over 2}
+\kappa_a\sqrt{{\ell(\ell+1)\over h(r_a)}-{1\over 4}} \;.
\label{26}
\end{equation}
Besides the surface gravity, the frequencies depend on only one other
parameter of the spacetime; this being the dimensionless parameter
$h(r_a)$.  On the basis of dimensional arguments, one might expect
$h(r_a)$ to be, at the very most, of the order unity and is, in fact,
exactly unity for the special case of Schwarzschild--de Sitter space.
[It is also worth pointing out that $h(r)$ depends on the mass
parameter, $m(r)$, but not on the redshift function, $\phi(r)$; that
is, the quasinormal modes, even at the level of the offset, cannot
give us a very detailed account of the dirty matter.]  Therefore, one
can see that, for any $\ell>0$, the argument of the square root is
positive and roughly proportional to $\ell^2$.  [Curiously, such
explicit $\ell$ dependence is notably absent in many studies on the
quasinormal modes for an {\it exterior} observer ({\it e.g.},
\cite{Nol,And,Motl,MN}). Nonetheless, there is no obvious
contradiction here, as no such work caters to our {\it
  squeezed}-horizon scenario.]

Meanwhile, $\ell=0$ requires special consideration, since the
square-bracket quantity in equation (\ref{22}) for the potential will
then vanish up to the order of $\Delta$.  If one goes on to examine
the next-order term in the brackets, it becomes evident that the
potential can no longer be cast into the Poschl-Teller form.
Nevertheless, it can be argued that this extra factor of $\Delta$ (for
$\ell=0$ the potential is of order $\Delta^3$ instead of $\Delta^2$)
causes the potential to become negligible.

It should be emphasized that, although approximations have been used
in attaining the Poschl--Teller form, equation (\ref{25}) provides an
{\it exact} description of the quasinormal mode behavior in the
squeezing limit.  This is because any neglected terms in the potential
are of, at most, the relative order $\Delta$ and
$\Delta\propto\kappa_a$.  Hence, we can more precisely write
\begin{equation}
k_{qnm}=\kappa_a \left[i\left(n+{1\over 2}\right)
+\sqrt{{\ell(\ell+1)\over h(r_a)}-{1\over 4}}
+{\cal O}[\Delta] \right].
\label{28}
\end{equation}
Moreover, the validity of this result does not depend on the horizons
becoming degenerate or even ``nearly degenerate''; they must only be
sufficiently close so that the condition $r_b-r_a\ll r_a$ is
satisfied.

Before concluding, let us briefly comment on the case of general $j$.
If one considers axial gravitational ($j=2$) perturbations, then the
scattering potential takes the form, as derived in the appendix, of
equation (\ref{pot}).  Closely following the prior methodology, we
find that, in the squeezed-horizon limit, this potential now becomes
\begin{equation}
V(r_*)\sim 
 {V_{2,\ell}\over \cosh^2\left[\kappa_a r_*\right]}
=  e^{-\phi(r_a)}{\kappa_a\left(r_b-r_a\right)\over
2\cosh^2\left[\kappa_a r_*\right]}{\left[\ell(\ell+1)-2\right]\over r^2_a}\;.
\label{29}
\end{equation}
Hence, all of the above outcomes carry through with the simple
replacement $V_{0,\ell}\rightarrow V_{2,\ell}\;$; that is,
$\ell(\ell+1)\rightarrow \ell(\ell+1)- 2\;$.  Although it is difficult
to be precise about general values of spin, one might be tempted to
suggest a ``generalization'' of the form $\ell(\ell+1)\rightarrow
\ell(\ell+1)+ F[j]\;$, where $F[j]$ is some simple polynomial of
leading order $j^2$.  Nevertheless, even though the $j$ dependence
probably deviates from such naive expectations, it is still clear that:
\\
---{\it (i)} the qualitative features of the analysis will persist, and,
\\
---{\it (ii)} for large enough values of $\ell$ the dependence on $j$
will become negligible.

\section{Conclusion}
\label{S:IV}

In summary, we have determined the asymptotic frequencies of the
quasinormal modes as would be measured by an observer who is trapped
between a pair of ``squeezed'' horizons.  Although these circumstances
are arguably special, our analysis is quite general in the sense that
the spacetime can contain arbitrary matter fields or ``dirt''.  That
is, we have allowed for the most general horizon geometries in a
static and spherically symmetric spacetime; thus extending the
Schwarzschild--de Sitter treatment of Cardoso and Lemos~\cite{CL}.

We have found that the ``gap'' (or asymptotic level spacing) coincides
precisely with the surface gravity, $\kappa_h$ (at either horizon); in
notable agreement with our other recent paper~\cite{MMV}. Moreover,
the ``offset'' (or subleading term) was found to go, roughly, as
$\kappa_h \ell$ for large enough $\ell$ and $n$.  Let us re-emphasize
that our results are {\it exact} in the squeezing limit, that is,
provided that the spacing between the horizons is small in units of
horizon radius.

We should also stress that the scenario considered here is distinct
from what is usually referred to, in the literature, as near
extremality or horizon degeneracy.  Studies on (nearly) extremal black
holes usually presuppose an observer who remains exterior to both of
the horizons.  One should not expect the outcomes of this paper to
translate into this other type of scenario. Indeed, in our related
work~\cite{MMV}, we have argued that such an exterior observer would
likely find the quasinormal modes to collapse, asymptotically, to zero
momentum.~\footnote{Actually, the extremal situation is somewhat more
  complicated than this. See~\cite{MMV} for a detailed discussion.}

Finally, let us ponder upon the meaning (if any) of an offset term
that depends so strongly on the orbital angular momentum, $\ell$, of
the perturbation field.  To remind the reader, the offset of a (for
instance) Schwarzschild black hole is believed to asymptote towards
the same fixed value [$\kappa \ln(3)/2\pi$] for any (fixed) choice of
$\ell$~\cite{Nol,And,Motl,MN}.  There is, of course, no contradiction
here; any scenario with an {\it exterior} observer is decidedly
different from ours.  Nonetheless, this ``discrepancy'' does raise
issues with a popular conjecture: the asymptotic value of the offset
can be employed to fix the spacing between black hole area
eigenvalues~\cite{Hod,Dre}.  If this were truly the case, then it
seems rather strange that the asymptotic value of the offset can, at
times, depend on the angular momentum of the particles being emitted.
That is to say, one would expect this $\ell$ independence to persist
for all types of black holes or, alternatively, why should only
certain classes of horizons be subject to quantization?  It would seem
that the status of using quasinormal modes, in this particular
context, requires further investigation.

\section*{Acknowledgments} 

Research supported by the Marsden Fund administered by the New Zealand
Royal Society and by the University Research Fund of Victoria
University.

\section*{Appendix: Wave equation for gravitational perturbations}
\label{A:A}

One can can always see how a gravitational field is perturbed by
simply perturbing the relevant metric.  Holding the stress tensor
fixed and studying the linear perturbations, one should then be able
to extract the gravitational wave equation and, thus, the scattering
potential for a graviton.  Note, however, that this technique says
nothing about {\it general} $j=2$ fields; that is, we will be strictly
considering $j=2$ perturbations due to gravity.  [In this regard, let
us point out that much of the quasinormal literature assumes
generalizations (from the scalar case) are straightforward, but (as
observed by Cveti\v{c} and Larsen, and by Kanti and
March--Russell~\cite{spin-papers}) such a naive approach fails to
incorporate the Bianchi identities except in the vacuum equations.
Which is to say, the wave equation should, generally, depend on the
{\it type} of field being discussed and not just its spin.]  The
computation reported below is an independent consistency check on that
of Karlovini~\cite{Karlovini}.

Allowing for the possibility of gravitational radiation, we will
(closely following Chandrasekhar~\cite{Chandra-book}) consider small
perturbations of the metric and characterize these by $q_r, q_t$ and
$q_\theta$:
\begin{eqnarray}
\d s^2 = -e^{-2\phi(r)}\left(1-\frac{2m(r)}{r}\right)\d t^2 +
 \left( 1- \frac{2m(r)}{r}\right)^{-1}\d r^2 + r^2 \d \theta^2 
\label{eq:DM:metric}\\
\quad\quad\quad + r^2\sin^2\theta\left[\d \varphi - q_r(r,\theta,t)\d r 
- q_\theta(r,\theta,t)d\theta - q_t(r,\theta,t)\d t\right]^2\;. \nonumber
\end{eqnarray}
(So far, all the functional dependence is explicit in the metric but,
for the sake of brevity, this will not always be so in the following.)
The above equation describes the so-called \emph{axial} perturbations,
for which a change in the sign of $\varphi$ necessitates that the
perturbations must also change sign (for the metric to remain
unchanged).  It should be emphasized that these perturbations are not
completely general; one could, just as well, always introduce small
perturbations into the functions $m(r)$ and $\phi(r)$.  This other
(neglected) type are called \emph{polar} perturbations; for this kind,
a reversal of $\varphi$ does not affect the metric. In the case of a
vacuum, axial and polar perturbations must decouple and any arbitrary
perturbation can be written as a linear superposition.  Let us note,
however, that even for the simple model of a Schwarzschild black hole
in a vacuum, the scattering potential for polar perturbations does
{\it not} generalize to the (naive) suggestion of replacing $2m$ with
$2m(1-j^2)$ in equation (\ref{11}) for scalars.

We now proceed by demanding that the perturbations do not change the
stress-energy tensor (at least) to first order in $q$. The
non-vanishing first-order components of the Ricci tensor are found to
be
\begin{eqnarray}
\delta R_{13}=\frac{1}{r^3\sqrt{1-2m/r}\sin^2\theta\;
e^{-\phi}}\left[q_{[r,\theta]}
r^2\sin^3\theta\left(1-\frac{2m}{r}\right)
e^{-\phi}\right]_{,\theta} \\
\quad\quad\quad\quad\quad\quad\quad\quad\quad\quad
+\frac{r \sin\theta}{e^{-2\phi}\sqrt{1-2m/r}}q_{[t,r],t}\;,   
\nonumber\\
\delta R_{23}=-(r^2\sin^2\theta
e^{-\phi})^{-1}\left[q_{[r,\theta]}
r^2\sin^3\theta\left(1-\frac{2m}{r}\right)
e^{-\phi}\right]_{,r} \\
\quad\quad\quad\quad\quad\quad\quad\quad\quad\quad
+ \frac{\sin\theta}{e^{-2\phi}(1-2m/r)}q_{[t,\theta],t}\;, 
\nonumber\\
\delta R_{03}=\frac{1}{r^3\sin^2\theta}
\left[\sqrt{1-2m/r}\left(q_{[t,r]}
\frac{r^4\sin^3\theta}{e^{-\phi}}\right)_{,r} \right.
\\ \quad\quad\quad\quad\quad\quad\quad\quad\quad\quad 
\left. 
+\frac{1}{re^{-\phi}\sqrt{1-2m/r}}
\left(q_{[t,\theta]}
r^3\sin^{3}\theta\right)_{,\theta}\right]\;, \nonumber
\end{eqnarray}
which yield the relations
\begin{eqnarray}
\delta R_{13} = 0 \Leftrightarrow 
\left[q_{[r,\theta]}r^2
\sin^3\theta\left(1-\frac{2m}{r}\right)
e^{-\phi}\right]_{,\theta} = -\frac{r^4\sin^3\theta}{e^{-\phi}}
q_{[t,r],t}\label{eq:peturb1}\;,
\\
\delta R_{23} = 0 \Leftrightarrow 
\left[q_{[t,\theta]}r^2\sin^3\theta
\left(1-\frac{2m}{r}\right)e^{-\phi}\right]_{,r} 
= \frac{r^2\sin^3\theta}{e^{-\phi}(1-2m/r)}
q_{[t,\theta],t}\label{eq:peturb2} \;,
\end{eqnarray}
with the third possible expression being redundant.  These equations
can be most easily found by tetrad methods or, alternatively, consult
\S 24 of Chandrasekhar~\cite{Chandra-book} [equations (11) and (12)].
Note that these formulas even hold true in the more general case of
$g_{tt}$, $g_{rr}$ and $g_{\theta\theta}$ depending on $\theta$ as
well.

To help simplify the above results, let us define the variable $Q$ by
\begin{equation}
Q(r,\theta,t) 
\equiv 
r^2\sin^3\theta\left[1-\frac{2m(r)}{r}\right]e^{-\phi(r)}q_{[r,\theta]}
\;. 
\end{equation}
The equations governing the perturbations [(\ref{eq:peturb1} and
\ref{eq:peturb2})] can now be expressed in the following form:
\begin{eqnarray}
Q_{,r} &= \frac{r^2\sin^{3}\theta}{e^{-\phi}(1-2m/r)}\;q_{[t,\theta],t}
\;,   \label{blah}\\
Q_{,\theta} &= -r^4\sin^3\theta \;e^{\phi}\; q_{[t,r],t} \; \label{blah2}.
\end{eqnarray}

Next, we can solve for separable harmonic perturbations by imposing a
suitable {\it ansatz},
\begin{equation}
Q(r,\theta,t) = rR(r)\;\vartheta(\theta)\; e^{i k t}\;,
\end{equation}
with other solutions obtainable by superposition.  Also applying the
commutativity property of the partial derivatives, we can eliminate
the function $q_t$ from equations (\ref{blah}) and (\ref{blah2}) to
obtain
\begin{eqnarray}
\vartheta\left[\frac{2e^{-\phi}
(1-2m/r)}{r^2\sin^3\theta}(rR)_{,r}\right]_{,r} 
+ \frac{R e^{-\phi}}{r^3}
\left[\frac{2 (\vartheta)_{,\theta}}{\sin^3\theta} 
\right]_{,\theta} \\
\quad\quad\quad\quad\quad\quad 
= k^2(q_{\theta,r}-q_{r,\theta})e^{-i k t} 
=  -\frac{2k^2 R\vartheta}{r\sin^3\theta(1-2m/r)e^{-\phi}}\;,
\nonumber
\end{eqnarray}
which implies 
\begin{eqnarray}
\frac{r^3 e^{\phi}}{R}
\left[\frac{e^{-\phi}(1-2m/r)}{r^2}(rR)_{,r}\right]_{,r} 
+ \frac{\sin^3(\theta)}{\vartheta}
\left[\frac{1}{\sin^3(\theta)} (\vartheta)_{,\theta}\right]_{,\theta} 
\\
\quad\quad\quad\quad\quad\quad
+ r^2 e^{2\phi}\frac{k^2}{(1-2m/r)} 
= 0\; \nonumber.
\end{eqnarray}
The terms are now in a form that is particularly suitable for finding
separable solutions. Denoting the separation constant by $\gamma^2$,
one finds that
\begin{eqnarray}
r^3 e^{\phi}\left[\frac{e^{-\phi}(1-2m/r)}{r^2}(rR)_{,r}\right]_{,r} 
+ r^2 e^{2\phi}\frac{k^2}{(1-2m/r)} R - \gamma^2 R &= 0\;,\label{blah3}\\
\sin^3\theta\left[\frac{1}{\sin^3\theta} 
\vartheta_{,\theta}\right]_{,\theta} + \gamma^2\vartheta &= 0\;.
\label{blah4}
\end{eqnarray}

The angular equation is quite easily solved and gives
\begin{equation}
\vartheta(\theta) = \sin^2\theta
\left[A \; P^{\ell}_{2}\left(\cos(\theta)\right) 
+ B \; Q^{\ell}_{2}\left(\cos(\theta)\right)\right]\;,
\end{equation}
where 
\begin{equation}
\ell\equiv{\sqrt{9+4\gamma^2}-1\over2},
\end{equation}
$P^{\ell}_m$ and $Q^{\ell}_{m}$ are the Legendre polynomials, while
$A$ and $B$ are arbitrary constants.  To ensure that the solutions are
real, we must restrict the values of $\ell$ to integers, and so
\begin{equation}
\gamma^2=\ell^2 + \ell - 2 = (\ell + 2)(\ell - 1)\;
\qquad {\rm where} \qquad \ell=0,1,2,...\; .
\end{equation}

The radial equation is not so trivially solved. However, we can still
transform it into a recognizable form by first taking note of equation
(\ref{2}) for the (generalized) tortoise coordinate.  It follows that
the radial equation can also be expressed as
\begin{equation}
r^3 e^\phi\left[\frac{\d r}{\d r_*}\frac{R}{r^2} 
+ \frac{R_{,r_*}}{r}\right]_{,r} 
+ r^2 e^{\phi}\frac{\d r_*}{\d r} k^2 R - \gamma^2 R = 0
\end{equation}
or
\begin{equation}
\frac{\d^2 R}{\d r_*^2} 
+ \frac{\d r}{\d r_*}
\left[{1\over r}\frac{\d\phantom{r}}{\d r}
\left(\frac{\d r}{\d r_*}\right)R 
- 2\left(\frac{\d r}{\d r_*}\right)\frac{R}{r^2} 
- \frac{\gamma^2 R}{r^2}e^{-\phi}\right] = -k^2 R \;.
\end{equation}

The above can readily be identified as having the form of a
one-dimensional scattering equation [{\it cf}, equation (\ref{10})]
with a potential of
\begin{equation}
V(r) \equiv \frac{\d r}{\d r_*}
\left[-{1\over r} \frac{\d\phantom{r}}{\d r}
\left(\frac{\d r}{\d r_*}\right) 
+ 2\left(\frac{\d r}{\d r_*}\right)\frac{1}{r^2} 
+ \frac{\gamma^2}{r^2}e^{-\phi}\right] \label{eek} 
\end{equation}
or
\begin{equation}
V(r)=e^{-2\phi}\left(1-\frac{2m}{r}\right)
\left[\frac{\ell(\ell + 1)}{r^2} - \frac{6m}{r^3}
+ \frac{2m_{,r}}{r^2} +
\left(1-\frac{2m}{r}\right)\frac{\phi_{,r}}{r} 
 \right]\;.
\label{pot}
\end{equation}
(As a consistency check, it can be verified that this form correctly
reduces to the potential given in~\cite{CL} for the same type of
perturbations in a Schwarzschild--de Sitter background.)  This
equation can be somewhat simplified by noting that the two derivative
terms appearing above can be rewritten in terms of the
$R_{\hat\theta\hat\theta}$ (or $R_{\hat\varphi\hat\varphi}$)
orthonormal component of the Ricci tensor:
\begin{equation}
V(r)=e^{-2\phi}\left(1-\frac{2m}{r}\right)
\left[\frac{\ell(\ell + 1)}{r^2} - \frac{6m(r)}{r^3}
+ R_{\hat\theta\hat\theta}(r)
 \right]\;.
\label{pot2}
\end{equation}
Equivalently, in terms of the Einstein tensor
\begin{equation}
V(r)=e^{-2\phi}\left(1-\frac{2m}{r}\right)
\left[\frac{\ell(\ell + 1)}{r^2} - \frac{6m(r)}{r^3}
+ {G_{\hat t\hat t} -  G_{\hat r\hat r}\over 2}
 \right]\;.
\label{pot3}
\end{equation}
If desired, the Einstein equations could the be used to rewrite the
trailing term in terms of the density and radial pressure.  Comparing
this potential with equation (\ref{12}) for a scalar perturbation, one
can see that there is no obvious generalization when the black hole is
truly ``dirty''.  As a further point of interest, one can also see
[most easily from equation (\ref{eek})] that, in the limit of squeezed
horizons, only the $\gamma^2$ term will contribute to the scattering.

As a second consistency check this can be compared with the analysis of
Karlovini~\cite{Karlovini}, to which it is equivalent for $j=2$. In
fact Karlovini also performed the $j=1$ calculation for the Maxwell
field. Combined with the (minimally coupled) scalar result as
discussed earlier in this article it is possible to re-cast Karlovini's
result in the form
\begin{equation}
V(r)=e^{-2\phi}\left(1-\frac{2m}{r}\right)
\left[\frac{\ell(\ell + 1)}{r^2} + (1-j^2) \frac{2m(r)}{r^3}
- (1-j) R_{\hat \theta\hat \theta} 
 \right]\;,
\label{pot4}
\end{equation}
for $j=0$, $1$, $2$. Given the results of the
scalar, Maxwell, and gravity computations, deriving this is a
curve-fitting exercise on a curve with 3 points.
There is therefore no particular reason to trust
this formula for higher values of $j$.

\section*{References}

\end{document}